\documentclass[12pt]{iopart}
\usepackage{graphicx}
\usepackage{dcolumn}
\usepackage{iopams}  
\usepackage{color}
\usepackage{bm}

\newcommand{\ket}[1]{\left|#1\right\rangle} 
\newcommand{\bra}[1]{\left\langle#1\right|} 
\newcommand{\op}[1]{\ensuremath{\hat{\mathrm{#1}}}}

\def\vec#1{{\boldsymbol{#1}}}  

\begin{document}

\title[FQH states of a Bose gas with spin-orbit coupling]{Fractional quantum 
Hall states of a Bose gas with spin-orbit coupling}

\author{T. Gra\ss$^1$, B. Juli\'a-D\'iaz$^1$, M. Burrello$^2$, and M.
Lewenstein$^{1,3}$} 

\address{$^1$ ICFO-Institut de Ci\`encies Fot\`oniques, Parc Mediterrani 
de la Tecnologia, 08860 Barcelona, Spain}

\address{$^2$ Instituut-Lorentz, Universiteit Leiden, P.O. Box 9506, 2300 RA
Leiden, The Netherlands}

\address{$^3$ ICREA-Instituci\'o Catalana de Recerca i Estudis
Avan\c cats, 
08010 Barcelona, Spain}
\ead{tobias.grass@icfo.es}

\begin{abstract}
We study the fractional quantum Hall phases of a 
pseudospin-1/2 Bose gas in an artificial gauge field. 
In addition to an external magnetic field, the gauge 
field also mimics an intrinsic spin-orbit coupling of the Rashba type.
While the spin degeneracy of the Landau levels is lifted by the spin-orbit
coupling, the crossing of two Landau levels at certain coupling strengths gives
rise to a new degeneracy. We therefore take into account two Landau levels, and
perform exact diagonalization of the many-body Hamiltonian. We study and
characterize the quantum Hall phases which occur in the vicinity of the
degeneracy point. Notably, we describe the different states appearing at the
Laughlin fillings, $\nu=1/2$ and $\nu=1/4$. While for these filling factors
incompressible phases disappear at the degeneracy point, denser systems at
$\nu=3/2$ and $\nu=2$ are found to be clearly gapped. For filling factors
$\nu=2/3$ and $\nu=4/3$, we discuss the connection of the exact ground state to
the non-Abelian spin singlet states, obtained as the ground state of $k+1$ body
contact interactions.
\end{abstract}

\vspace{2pc}
\maketitle

\section{Introduction}

Cold atoms provide versatile and highly controllable quantum 
systems. This makes them particularly useful as quantum simulators, 
with plenty of applications which stem from their complicated 
many-body behavior. Cold atoms have been used to explore important open problems
in solid state physics. A noteworthy example is the physics emerging
from strongly correlated electronic systems, which has fascinated
researchers for decades. Due to its topological nature and the
exotic anyonic excitations, fractional quantum Hall (FQH) systems 
are particularly interesting candidates for a quantum simulation 
with cold atoms~\cite{laughlin, arovas, moore-read, wen-niu}. 

A major difference between electrons and atoms, however, is 
the atoms' lack of electric charge. This makes them 
insensitive to external magnetic fields, which would hinder 
exploring the vast field of quantum Hall physics. Different 
proposals to overcome this have been made: In principle, 
rotating the atomic system is sufficient to mimic the 
Lorentz force by mean of the Coriolis force~\cite{cooper-aip,fetter-revmod} 
and bring the system into quantum Hall 
phases~\cite{cooperwilkin,wilkingunn,cooper-wilkin-gunn}. 
However, experimental attempts to reach this regime by 
rotating the cloud have not been entirely 
successful~\cite{schweikhard,cornell-vortices,cornell-vortex-lattice,dalibard-vortex2002,dalibard-vortex2004}. 
Alternatively, the magnetic field can be mimicked 
by dressing the atoms with laser fields in such a way that 
the lowest dressed state obtains a space-dependent phase factor. 
Properly chosen, this phase factor acts as if the atoms 
were moving like charged particles in a magnetic 
field~\cite{dalibard,brunoPRA,bruno-njp}. A Raman dressing 
of $^{87}$Rb atoms has been used to generate an artificial 
electric field~\cite{spielmanPRL}. Later on, this scheme 
has successfully mimicked an artificial magnetic field 
and visibly induced vortices into the atomic cloud~\cite{lin}. 
Artificial gauge fields have also been realized in optical 
lattices, by means of laser dressing~\cite{spielman-peierls, bloch-gauge} 
or shaking of the lattice \cite{sengstock12}.

These schemes allow for simulating the behavior of charged 
spinless (or spin-polarized) particles. The spin degree 
of freedom of electrons might play a role in fractional 
quantum Hall setups, as the strong magnetic field does 
not necessarily fully polarize the electron spin due to 
the small gyromagnetic ratio in many solid materials. 
Allowing the atoms to be in two internal states, one might 
try to mimic spin-1/2 particles like electrons, but again 
the lack of charge will cause significant differences  
between this pseudospin, and the real spin of the electrons. 
The latter interacts with a magnetic field which is induced 
by the motion of the charged electron through the electric 
field of the nuclei. This effect, called Rashba spin-orbit 
coupling~\cite{rashba}, is absent in cold atoms. Ways to 
overcome this have been proposed recently. The laser 
dressing of atoms can be designed in such a way that it 
not only implements the effect of a magnetic field, but 
also of a spin-orbit coupling~\cite{juzeli2004,fleischhauer,galitski}. 
Mathematically, spin-orbit coupling and magnetic fields 
can be treated on the same footing: While the magnetic 
field can be taken into account through a minimally coupled 
vector potential $\vec{A}$, i.e. a term $\vec{p}\cdot\vec{A}$ 
in the Hamiltonian, the spin-orbit coupling has the form 
$\vec{p}\cdot\vec{\sigma}$, where 
$\vec{\sigma}=(\sigma_x,\sigma_y,\sigma_z)^T$, 
a vector of Pauli matrices. This makes the spin-orbit 
coupling equivalent to gauge fields, which in the case 
of spin-1/2 belong to the SU(2) group, and are thus 
in general referred to as \textit{non-Abelian} gauge fields. 
In pioneering experiments, spin-orbit coupling has been 
synthesized experimentally for $^{87}$Rb~\cite{spielman-sobec,williams12}, 
using a Raman dressing similar to the one of Refs.~\cite{spielmanPRL,lin}. 
These experiments have attracted a great deal of attention, 
and the impact of spin-orbit coupling on the condensation 
properties of a Bose gas has been studied 
theoretically~\cite{sinha11,hu12,ryan,ryan-erratum,ozawa-baym-stability,ozawa-baym-stripes,SOstringari}. 
Proposals to synthesize spin-orbit coupling for atoms in 
optical lattices~\cite{osterloh,philipna} also motivate 
the study of Bose-Hubbard models with spin-orbit 
coupling~\cite{indianpra,trivediBH,galitski12}.

A different scenario is encountered in the presence of 
both a spin-orbit coupling and a strong artificial magnetic 
field. The latter induces strong correlations, producing 
fractional quantum Hall phases which in turn may be qualitatively 
affected by the spin-orbit coupling. Its influence could 
perfectly be revealed by a cold atom experiment which 
combines the artificial spin-orbit coupling and the 
artificial magnetic field, and allows to freely tune 
the coupling strength. The closest connection to the electronic 
fractional quantum Hall effect can be made by considering 
fermionic atoms with dipolar interactions~\cite{fqhe-gap}. 

In this paper, we consider the experimentally more 
feasible scenario of bosonic atoms with short-range 
interactions. They are confined to a plane, or, 
as we will assume periodic boundary conditions, to a torus. The bosons are
subjected to an artificial magnetic 
field perpendicular to the plane, and we assume a 
Rashba coupling between an effective pseudospin-1/2 
degree of freedom and the external motion of the atoms. 
It has recently been shown that in the absence of 
spin-orbit coupling but within an artificial magnetic 
field a two-component Bose gas forms spin 
singlets~\cite{rapido,furukawa}. The strongly correlated 
regime is then described by a series of states with 
non-Abelian anyon excitations (non-Abelian spin singlet or NASS states)~\cite{ardonne-schoutens,nass-nucl}. 
It has also been shown that via a Rashba-type spin-orbit 
coupling a certain spin polarization is favored, resulting 
in spin polarized quantum Hall phases~\cite{trombettoni,trombettoniPRA} 
which are derived from the Read-Rezayi 
series~\cite{read-rezayi}. Distinct quantum Hall phases, 
however, have been predicted for particular spin-orbit 
coupling strengths which correspond to a degeneracy between 
two Landau levels~\cite{trombettoni,trombettoniPRA}. 
Numerical studies restricted to a single Landau level 
have been able to back the occurrence of the 
Read-Rezayi-like phases~\cite{palmer-pachos,komineas-cooper}. 
For the interesting configuration at the degeneracy points,
gapped phases at the NASS filling factors have been numerically proven 
recently~\cite{rapido}.

We undergo a complete numerical study, 
complementing the results presented in~\cite{rapido}, thus 
taking into account two Landau levels. Performing 
exact diagonalization with such an increased basis, our 
study remains valid also at the degeneracy points. We 
start with a detailed description of the system in 
Sec.~\ref{system}. Next, we solve the single-particle problem 
in Sec.~\ref{LLstruc}. Afterward, we turn to the 
many-body problem in Sec.~\ref{mb}, calculating the 
interaction matrix elements. With this we can perform 
the exact diagonalization, yielding the results presented 
in Secs.~\ref{laugfate} and~\ref{secim}. Our main findings
are summarized in Sec.~\ref{sum}.

\section{The system \label{system}}

We consider a pseudospin-1/2 Bose gas in two dimensions, 
described by a single-particle Hamiltonian 
\begin{eqnarray}
\label{Hi}
H_i = \frac{1}{2M} \left[\vec{p}_i {\bf 1}_2 +\vec{A}(\vec{r}_i)\right]^2,
\end{eqnarray}
with ${\bf 1}_2$ being the $2\times2$ identity matrix. 
An artificial gauge potential $\vec{A}$ mimics a constant, 
perpendicular magnetic field and a coupling between the 
pseudospin and the external motion. For convenience, we 
choose this coupling to be of the Rashba-type. In the 
Landau gauge, the vector potential reads
\begin{eqnarray}
\label{Asp}
 \vec{A} = B (0,x,0) {\bf 1}_2  + q (\sigma_x,\sigma_y,0).
\end{eqnarray}
Here, $B$ is the magnetic field strength, $q$ controls the 
strength of the spin-orbit coupling, and $\sigma_i$ are 
Pauli matrices. Both contributions to this gauge potential 
have already been engineered experimentally separately: An artificial 
magnetic field of this form (i.e. $q=0$) has been realized 
through an atom-laser dressing in Bose-Einstein condensates 
without pseudospin degree of freedom~\cite{lin}, and 
through rotation in two-component 
systems~\cite{cornell-vortices,cornell-vortex-lattice}. A similar 
spin-orbit coupling has been achieved experimentally by a 
Raman coupling of three atomic states with two 
lasers~\cite{spielman-sobec,williams12}. A theory proposal of 
how to achieve a gauge potential combining both the magnetic 
and the spin-orbit part has been given in 
Ref.~\cite{trombettoni, trombettoniPRA}. Here, the spin-orbit 
term arises due to a tripod coupling of three atomic ground 
states to one excited state by three laser fields. A review 
of different proposals is given in Ref.~\cite{dalibard}.

All schemes have in common that they effectively consider 
dressed atoms, that is superpositions of different bare 
atomic states. It is in the basis of dressed states that 
the single-particle Hamiltonian takes the desired form of 
Eq.~(\ref{Hi}). While equivalent on the single-particle 
level, the different proposals might require different 
descriptions for the particles' interactions. Tripod 
schemes with different laser configurations, though 
mimicking the same spin-orbit coupling, have been shown 
to support different mean-field phases due to differences 
in the interaction term~\cite{tripod-interactions}.
It has also experimentally been shown that the interactions 
between dressed atoms might become long-range, though 
the bare atoms interact via $s$-wave scattering~\cite{williams12}. 
Also if the interaction potential of the dressed atoms 
effectively remains a contact potential, the scattering 
lengths may be modified. In the first experimental realization 
of spin-orbit coupling, the strength of interactions between 
different pseudospin states behaved linearly with the 
effective spin-orbit coupling strength, which caused a 
quantum phase transition between a pseudospin-mixed phase 
and a pseudospin-separated phase upon tuning the spin-orbit 
coupling strength~\cite{spielman-sobec}.

Thus, the choice of interactions is a delicate issue. In 
this paper, we consider two-body contact interactions 
between the dressed atoms. 
\begin{eqnarray}
\label{V}
 V(\vec{r}_i,\vec{r}_j) = \frac{\hbar^2}{M} \delta^{(2)}(\vec{r}_i-\vec{r}_j) 
\left(
\begin{array}{cc}
 g_{\uparrow\uparrow} & g_{\uparrow\downarrow} \\
 g_{\uparrow\downarrow} & g_{\downarrow\downarrow}
\end{array}
\right)\,.
\end{eqnarray}
Such a description is appropriate for the pioneering experiment in Ref.~\cite{spielman-sobec}. In this experiment, $^{87}$Rb has been used with nearly equal scattering lengths in and between different hyperfine states. Also, for weak or moderate spin-orbit coupling strength the scheme of 
Ref.~\cite{spielman-sobec} does not significantly alter the effective scattering lengths for the dressed atoms. It is thus convenient to assume SU(2) symmetry, that is, $g_{\uparrow\uparrow}=g_{\downarrow\downarrow}=g_{\uparrow\downarrow} \equiv g$. 

Previous studies have performed numerical diagonalization 
of pseudospin-1/2 systems without spin-orbit coupling 
($q=0$)~\cite{rapido,furukawa}. We note that in this case, 
the pseudospin polarization $S$ defined as the difference 
$N_{\uparrow}-N_{\downarrow}$ of up- and down-spin particles is 
a conserved quantity. This reduces the size of the Hilbert 
space, and yields eigenstates of well-defined polarization. For SU(2)-symmetric
interactions, at almost all filling factors the ground state is found to be a
pseudospin singlet, $S=0$. For bosons interacting via a two-body contact 
potential, eigenstates with zero interaction energy exist 
up to a filling factor $\nu=2/3$. Here the (221) Halperin 
state becomes the unique zero interaction energy state, if one does not
take into account edge excitations, as is the case within our numerics
performed on a torus. The wave function of the (221) Halperin state on a
disk reads~\cite{halperin}
\begin{eqnarray} 
\label{221-halperin}
\Psi_{\mathrm{H}} = \prod_{i<j}^{N_{\uparrow}} (z_{i \uparrow}-z_{j \uparrow})^2
\prod_{k<l}^{N_{\downarrow}} (z_{k \downarrow}-z_{l \downarrow})^2 \prod_{i,k}
(z_{i \uparrow} - z_{k \downarrow}) {\rm e}^{-\sum \frac{|z_i|^2}{4}}\,,
\end{eqnarray}
where we have introduced the complex variables, $z_j=x_j-i y_j$.

This Halperin state is part of a series of states occurring at 
filling factors $\nu=2k/3$, defined as the maximum-filled 
zero-energy eigenstates of a $(k+1)$-body contact interaction. 
This series is commonly known as the NASS series, since the states with $k \geq 2$ support non-Abelian 
anyonic excitations~\cite{ardonne-schoutens,nass-nucl}. As 
pointed out in Refs.~\cite{rapido,furukawa}, the non-Abelian phase 
at filling $\nu=4/3$ is realized even in systems with a two-body 
contact potential.

Also systems with finite $q$ have been considered 
numerically~\cite{palmer-pachos,komineas-cooper}. The 
lowest Landau level approximation which has been made in 
these studies is valid within broad ranges of $q$ with 
sufficiently large Landau level gaps. These studies have 
shown that in these regimes the spin-polarized fractional 
quantum Hall states like the Laughlin state~\cite{laughlin} 
or Moore-Read state~\cite{moore-read} are relevant, if 
one maps them into the Landau level structure of the 
spin-orbit coupled system according to Eq.~(\ref{p0}). 
We will demonstrate this in more detail in Secs.~\ref{laugfate} 
and~\ref{secim} for the bosonic 
Laughlin state at filling factor $\nu=1/2$, investigating  
its fate when passing through the degeneracy point $q^2=3B$, and 
for other relevant filling factors which may present 
incompressible phases, respectively.

\section{Single particle Hamiltonian: Landau level structure \label{LLstruc}}

Prior to solving the many-body problem by numerical diagonalization, 
it is useful to solve the single-particle Hamiltonian. For the 
spin-orbit coupling of Eq.~(\ref{Asp}), this has been done 
in Refs.~\cite{trombettoni,trombettoniPRA},
yielding a Landau level structure which is simply a combination of two Landau
levels from the quantum 
Hall system without spin-orbit coupling. 
While they consider a system on a disk, for 
our numerical study it is advantageous to work in an edge-less geometry, 
as it allows to study bulk properties even for small system 
sizes. We thus choose a rectangle of sizes $a$ and $b$ with 
periodic boundary conditions, that is a torus, and perform the 
analog derivation for the Landau level structure.

For convenience, we choose units such that $\hbar=1$ and $M=1/2$, 
and define the following ladder operators:
\begin{eqnarray}
\label{cs}
 c &\equiv& \frac{1}{\sqrt{2B}} (p_x-ip_y -i B x), \\
 c^\dagger &\equiv& \frac{1}{\sqrt{2B}} (p_x+ip_y +i B x),
\end{eqnarray}
with $[c,c^{\dagger}]=1$ and $[c,c]=0$. The single-particle Hamiltonian 
can now be expressed in terms of these operators
\begin{eqnarray}
\label{Hsp}
H_{i} &=& 
\left(
 \begin{array}{cc}
2 B (c_i^{\dagger}c_i + \frac{1}{2}) + 2 q^2
& 
2\sqrt{2B} q c_i
\\
2\sqrt{2B} q c_i^{\dagger}
& 
2 B (c_i^{\dagger}c_i + \frac{1}{2}) + 2q^2
\\
\end{array}
\right)\,.
\end{eqnarray}
The diagonal part has the form of a harmonic oscillator, 
yielding the standard Landau levels 
$\tilde \Psi_{n,\kappa}(x,y) = \exp(i \kappa y) \ \Phi_{n,\kappa}(x)$
and 
\begin{equation}
\Phi_{n, \kappa}(x) = e^{-B ( x + \kappa/B )^2/2}\ \ H_n[\sqrt{B} (x + \kappa/B)]
\end{equation}
with $H_n$ the Hermite polynomials. The energy quantum number $n$ denotes 
the Landau level. Applying the periodic boundary conditions, as done 
in Ref.~\cite{yoshioka-halperin-lee}, $\kappa$ is restricted 
to integer multiples of $2\pi/b$. This yields a second quantum 
number $j \in \mathbb{Z}$ via the definition 
$X_j \equiv \kappa/B = 2\pi j / (b B)$. The quantity $X_j$ can 
be interpreted as a displacement of the harmonic oscillator in 
the $x$-direction. To ensure the periodicity of the wave function 
in $x$, with period $a$, we can sum over all displacements 
$X_j + k a$ with $k$ an integer, 
$\Psi_{n,j}(x,y) \equiv \sum_{k=-\infty}^{\infty} \tilde \Psi_{n,B(X_j + k a)}(x,y)$.
This will not affect the $b$-periodicity in $y$, if 
$ ab = 2\pi \lambda^2 N_{\Phi} $ with $N_{\Phi}$ an integer. To prevent 
double-counting, we must restrict $j$ to $1\leq j \leq N_{\Phi}$. 
We find that $N_{\Phi}$ quantizes the magnetic field, and can 
thus be interpreted the number of magnetic fluxes within the unit 
cell. It is connected to the particle number $N$ through the filling factor
\begin{eqnarray}
\label{ff}
 \nu &=& \frac{N}{N_{\Phi}} \equiv \frac{P}{Q},
\end{eqnarray}
where we also have introduced the co-prime integers 
$P$ and $Q$. Introducing the unitless variables 
$x'=x/\lambda$ and $y'=y/\lambda$, with $\lambda=1/\sqrt{B}$, 
we finally arrive at the wave functions of 
Ref.~\cite{yoshioka-halperin-lee}. In the lowest Landau 
level (i.e. for $n=0$), they read
\begin{eqnarray}
\label{sp2}
 \Psi_{0,j} &= & \left( 2 \pi^2 N_{\Phi}^3 \frac{a}{b} \right)^{-1/4}
{\rm e}^{ -\frac{y'^2}{2} - i x' y'} 
\ \theta_3 \left[ \frac{j\pi}{N_{\Phi}}
+\sqrt{\frac{\pi}{2N_{\Phi}a/b}}z, e^{-\frac{\pi}{N_{\Phi}a/b}} \right].
\end{eqnarray}
Here, $\theta_3$ is an elliptic $\theta$-function. The excited 
states can easily be obtained by applying the ladder operator 
$c^{\dagger}$, $\sqrt{n+1} \Psi_{n+1,j} = c^{\dagger} \Psi_{n,j}$.

Without spin-orbit coupling, i.e. for $q=0$ in Eq.~(\ref{Hsp}), 
the wave function 
\begin{eqnarray}
\label{ALLL}
 \vec{\Psi}_{0,j}^{q=0} \equiv \Psi_{0,j} (\alpha \ket{\uparrow} + \beta \ket{\downarrow}).
\end{eqnarray}
with $|\alpha|^2 + |\beta|^2 =1$ describes the lowest Landau 
level. Thus, apart from the degeneracy in $j$, we have an SU(2) 
degeneracy due to the pseudospin-1/2 degree of freedom.

\begin{figure}[t]
\flushright
\includegraphics[width=0.88\textwidth,angle=-0]{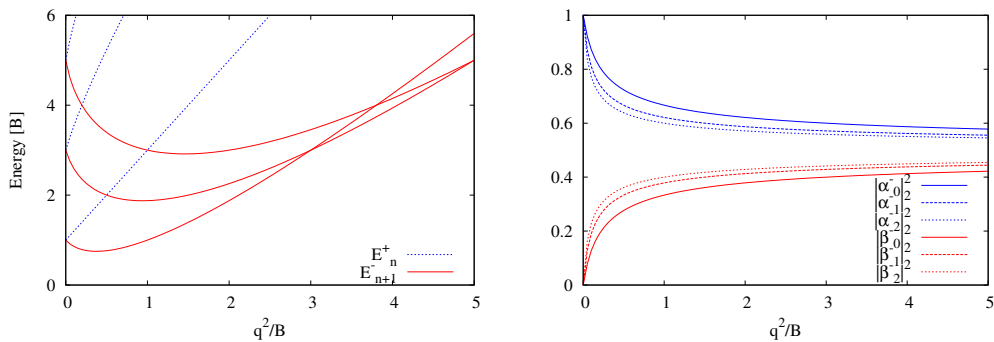}
\caption{\label{fig:ene} 
{\bf Left:} First eigenvalues of the single particle 
Hamiltonian, $E_n^\pm$ as a function of $q^2/B$, see Eq.~(\ref{spen}).
{\bf Right:} Weights of the two components of each eigenvector 
of the single particle Hamiltonian for the ``-'' states, $\alpha_n^-$, and 
$\beta_n^-$ as a function of $q^2/B$, see Eq.~(\ref{alphabeta}). }
\end{figure}

For finite $q$, we have to take into account higher Landau 
levels, as the off-diagonal elements in Eq.~(\ref{Hsp}) provide a Jaynes-Cummings-type coupling. Thus, the eigenspinors 
must be of the form 
\begin{eqnarray}
\label{eigenspinors}
\vec{\Psi}_{n,j}^{\pm} \equiv 
(\alpha_{n}^{\pm} \Psi_{n,j}, \beta_{n}^{\pm} \Psi_{n+1,j})^T,
\end{eqnarray}
where $\alpha_{n}^{\pm}$ and $\beta_{n}^{\pm}$ are functions 
of $q$. The single-particle spectrum is found to be the same as 
in Refs.~\cite{trombettoni,trombettoniPRA}, as expected due to 
gauge invariance. It reads
\begin{eqnarray}
\label{spen}
 E_n^{\pm} = 2B(n+1) +2q^2 \pm \sqrt{B^2+8Bq^2(n+1)}\,.
\end{eqnarray}
In Fig.~\ref{fig:ene} we present the lowest eigenvalues $E_n^\pm$. 

Since the eigensolutions corresponding to $E_{n}^+$ are higher 
than the lowest $E_{n}^-$ solutions, we can restrict ourselves 
to the latter. Their coefficients $\alpha_n^-$ and $\beta_n^-$ read
\begin{eqnarray}
\alpha_n^- &=& {\cal N} \left( B+2q\sqrt{2B(n+1)} + \sqrt{B^2+8 B q^2(n+1)} \right), \\
\beta_n^-  &=& {\cal N} \left( B-2q\sqrt{2B(n+1)} - \sqrt{B^2 +8 B q^2(n+1)} \right), \\
{\cal N}  &=& \frac{1}{2} \Big( B^2 + 8Bq^2(n+1)   + 2 q \sqrt{2 B(n+1)} \nonumber \\ 
& \times& \sqrt{B^2+8 B q^2(n+1)} \Big)^{-1/2}. \label{alphabeta}
\end{eqnarray}
In Fig.~\ref{fig:ene} we plot the value of the coefficients 
$\alpha^-_n$ and $\beta^-_n$ as a function of the spin-orbit coupling strength, 
$q$. The lower spin component quickly acquires weight, measured by $|\beta|^2$, as the spin-orbit coupling is turned on. 

We shall distinguish two cases, as seen in Fig.~\ref{fig:ene}, 
or Eq.~(\ref{spen}): 1) For broad ranges of $q$, the ground 
state is given by a single $E_n^-$ level. Thus, the pseudospin 
degeneracy of the system with $q=0$ is lifted, and the system 
is effectively spin-polarized. 2) The second case is restricted 
to separate points at $q^2/B = 2n+3$, where two solutions, 
$E_n^-$ and $E_{n+1}^-$, become degenerate. This scenario, with 
a two-fold degenerate single-particle ground state, is formally 
similar to the spin-degenerate case of Eq.~(\ref{ALLL}). We can 
define a pseudospin-1/2 spinor as, 
\begin{equation}
 \left(
\begin{array}{c}
 \gamma \\
 \delta
\end{array}
\right)
\equiv 
\gamma \, \vec{\Psi}_{0,j}^{-} 
+ \delta \, \vec{\Psi}_{1,j}^{-}  \qquad {\rm with}\; |\gamma|^2+|\delta|^2=1\,.
\end{equation}
The connection between these two Landau levels of the 
spin-orbit coupled system and the $q=0$, spin-degenerate 
lowest Landau levels, is made through the bijective mapping 
\begin{eqnarray}
\label{p0} 
 \ket{0,j,\uparrow} \mapsto \vec{\Psi}_{0,j}^{-} \equiv
\alpha_{0}^{-} \ket{0,j,\uparrow} + \beta_{0}^{-} \ket{1,j,\downarrow} 
\\ \label{p1}
 \ket{0,j,\downarrow} \mapsto \vec{\Psi}_{1,j}^{-} \equiv
 \alpha_{1}^{-} \ket{1,j,\uparrow} + \beta_{1}^{-} \ket{2,j,\downarrow} \,.
\end{eqnarray}
For brevity, in the following we denote these two levels 
by $\ket{n=0,j}$ and $\ket{n=1,j}$. In our numerical 
calculations, we will explicitly take into account these two levels, as we 
focus on configurations where $q^2/B$ is close to the first 
critical value of 3.  

\section{Many-body Hamiltonian\label{mb}}

From the single-particle states, we easily construct many-body Fock 
states 
\begin{equation}
\ket{n_{n=0,j=1},\cdots, n_{n=0,j=N_{\Phi}}, n_{n=1,j=1},\cdots,n_{n=1,j=N_{\Phi}}},
\end{equation}
and we can express the Hamiltonian in terms of annihilation 
and creation operators, $\op{b}^\dagger_{nj}$ and $\op{b}_{nj}$. 
These obey bosonic commutation relations and act on the  
occupation number $n_{nj}$ in the usual way. 
The Hamiltonian reads in this notation:
\begin{eqnarray}
 \label{Hcal}
 {\cal H} = 
\sum_{n,j} E_n^- \ \op{b}_{nj}^{\dagger} \op{b}_{nj} 
+ \sum_{\{n,j\}}  V_{\{n,j\}} \op{b}_{n_1j_1}^{\dagger} \op{b}_{n_2j_2}^{\dagger} \op{b}_{n_3j_3}\op{b}_{n_4j_4}.
\end{eqnarray}
The interaction matrix elements $V_{\{n,j\}}$, 
\begin{eqnarray}
 V_{\{ n,j \} } = \bra{n_{1},j_{1}} \bra{n_{2},j_{2}}  V \ket{n_{3},j_{3}} \ket{n_{4},j_{4}},
\end{eqnarray}
with $V$ given in Eq.~(\ref{V}), and $\{n,j\}$ 
denoting the set of quantum numbers $n_1,\dots,n_4$ 
and $j_1,\dots,j_4$, are given in~\ref{appA}.

Before numerically diagonalizing the Hamiltonian ${\cal H}$ of 
Eq.~(\ref{Hcal}), we shall exploit the translational symmetry 
of the system on a torus. This symmetry, fully discovered by 
F. D. M. Haldane~\cite{haldane}, yields a many-body basis where 
states are characterized by a ``Haldane momentum'' 
$\vec{K}=(K_x,K_y)$ which is conserved by the Hamiltonian. A 
comprehensive recipe for constructing this basis can be found in 
Ref.~\cite{chakraborty}. The main idea is to divide the many-body 
states into equivalence classes: An $N$-particle state characterized 
by the quantum numbers $\{j_1, \dots, j_N\}$ (and the $n$ quantum 
numbers which here do not matter) can by translation along $x$ 
be transformed into states with $\{j_1-qm, \cdots, j_N-qm\}$, 
where $q$ is defined in Eq.~(\ref{ff}), and $m$ is an integer. 
It runs from 0 to some maximum value $c_\eta$, where 
$ \{j_1, \dots, j_N\} = \{j_1-c_\eta m, \cdots, j_N-c_\eta m\}$ due 
to the definition of $j$ modulo $N_{\Phi}$. This set of $c_\eta$ 
different states forms an equivalence class which we label by $\eta$. 
We denote each state by $\ket{\eta, m}$. For each class $\eta$, 
the translational symmetric eigenstates are found as superpositions 
which can be characterized a quantum numbers $\tilde J$ running from 0 to
$N_\Phi/Q-1$:
\begin{eqnarray}
\label{newbasis}
 \ket{\eta,(\tilde J,J)} \equiv 
\frac{1}{\sqrt{c_\eta}} \sum_{k=0}^{c_\eta-1} 
\exp\left(\frac{2\pi i \tilde J k}{c_\eta}\right) \ket{\eta,k}.
\end{eqnarray}
A second quantum number $J$ is given by $J \equiv j_i \ \rm{mod} \ N_\Phi$. Both
quantum numbers $\tilde J$ and $J$ can be related to the 
Haldane pseudomomentum $\vec{K}$:
\begin{eqnarray}
\label{K-haldane}
\vec{K}\lambda =  
\sqrt{\frac{2\pi b}{N_{\Phi}a}} 
\left(\tilde J-\tilde J_0,\frac{a}{b}(J-J_0) \right).
\end{eqnarray}
The quantum numbers of zero pseudomomentum are defined as the 
point in the Brillouin zone with highest symmetry. In most 
cases,  $(\tilde J_0,J_0) = (0,0)$, but if $N$ is even and $P$ 
and $Q$ are both odd, other values become possible 
(cf. Ref.~\cite{chakraborty}). The numerical diagonalization 
is then performed within a Hilbert space with fixed filling 
factor $\nu$ and fixed pseudomomentum $\vec{K}$.

\section{The fate of the Laughlin state at $\nu=1/2$ \label{laugfate}}

\begin{figure}[t]
\vspace{20pt}
\flushright
\includegraphics[width=0.88\textwidth]{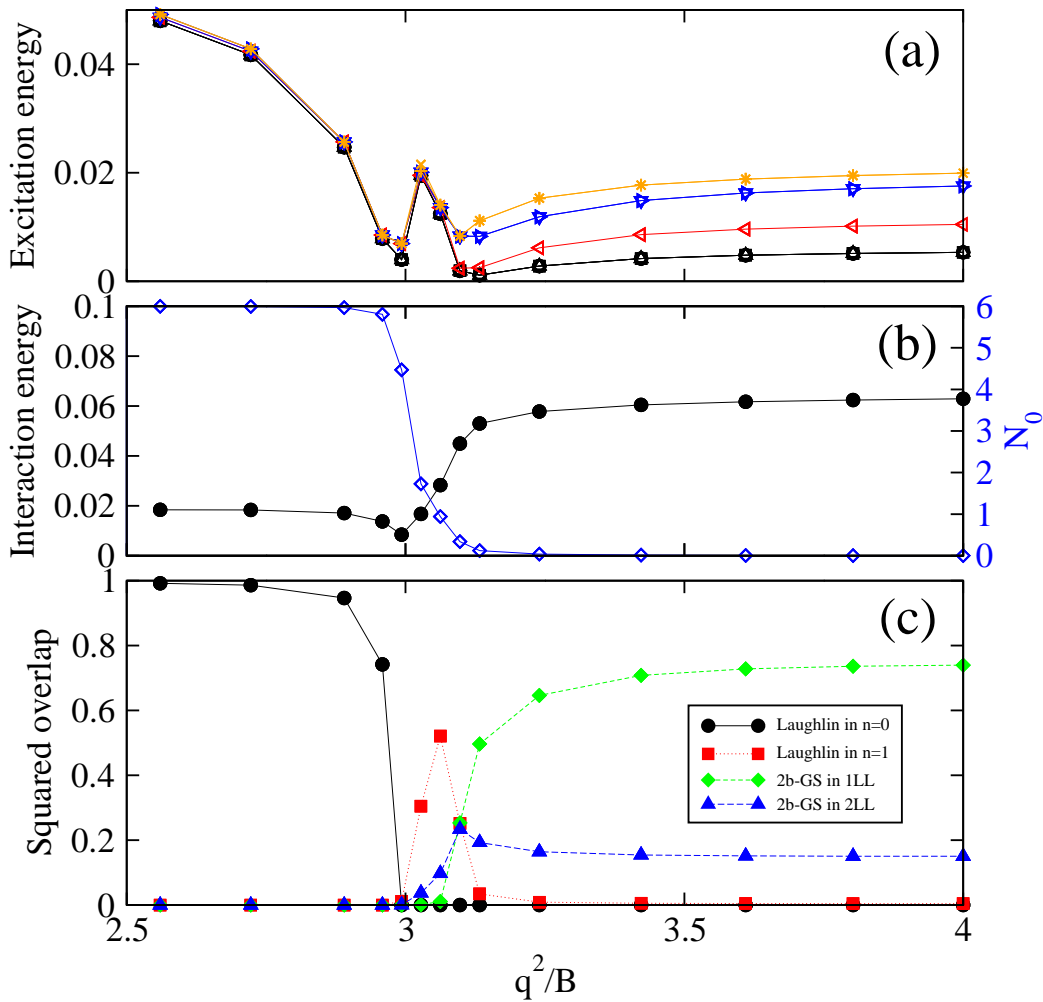}
\caption{\label{ovGS} Properties of the system 
as a function of the strength of the spin-orbit interaction, $q^2/B$.
{\bf (a)} Low-lying excitation energies. 
{\bf (b)} (left scale) Interaction energy of the ground 
state. (right scale) Occupation of $n=0$ level in the ground state.
{\bf (c)} Overlap of ground state with different test states 
as further explained in the text: Laughlin projections, and 
ground states of two-body contact interaction in the first 
and second excited LL. The system size is $N=6$ at half 
filling, with spin-independent interactions $g_{s_1s_2}=0.2$. 
Energies are in units of $gB$.}
\end{figure}

The Laughlin state at filling $\nu=1/Q$ is described on a disk by a wave
function:%
\begin{equation}
\label{laug}
\Psi_{\rm L}=
\prod_{i<j} (z_i-z_j)^Q  {\rm e}^{-\sum_i |z_i|^2/4} \,,
\end{equation} 
For $Q=2$, it constitutes the highest density eigenstate of a spinless or
spin-polarized system of bosons interacting via a two-body contact
potential, since its wave function becomes zero whenever two particles
coincide. We can thus easily find its second-quantized expression 
by exact diagonalization. Although performing the
mapping into the spin-orbit coupled Landau levels according to Eqs.~(\ref{p0})
and~(\ref{p1}) corresponds to simply redefining the basis of the Hilbert
space, this process has non-trivial physical implications. Since the
Laughlin state is a spinless state there is an arbitrariness 
in applying the mapping of either Eq.~(\ref{p0}) into the first or
Eq.~(\ref{p1}) into the second Landau level. The effect of this mapping to
higher Landau levels induces a sort of effective long range interaction
\cite{komineas-cooper} such that the $nu=1/2$ Laughlin state acquires in general
a
positive interaction energy. This is due to the fact that in the spin-orbit
coupled Landau level the corresponding wave function does not vanish when $z_i
\to z_j$.

Let us first investigate in which regimes of the spin-orbit coupling 
strength such projected Laughlin states describe well the ground 
state at filling factor $\nu=1/2$. Afterwards, we will consider Laughlin
states at smaller filling factors, which remain states of zero interaction
energy in the spin-orbit coupled system.

\subsection{Below the first degeneracy point: $0<q^2/B<3$}

For $q^2<3B$, the levels $\ket{n=0,j}$ are the lowest single-particle 
states, see Fig.~\ref{fig:ene}, thus we expect the ground state to 
be given by the projection into this manifold, following Eq.~(\ref{p0}). 
As shown in Fig.~\ref{ovGS} (c), the overlap of the 
true ground state with this projected Laughlin state is approximately 
1, if we are sufficiently far below the degeneracy point. However, 
it quickly decreases to 0 when approaching the degeneracy point. 
Simultaneously, the $n=0$ Landau level is depleted, see Fig.~\ref{ovGS} (b).

In Fig.~\ref{ovGS} (a), we plot the low-lying excitation 
energies as a function of $q^2/B$. This shows that the projected 
Laughlin phase is gapped. The gap, however, vanishes when approaching 
the degeneracy point. In Fig.~\ref{ovGS} (b), the 
interaction energy of the ground state is plotted. The projected 
Laughlin state has a finite interaction energy, due to interactions 
within the lower component and between the components. A sudden 
decrease in the interaction energy is observed when the degeneracy 
point is approached.

The numerical results shown in Fig.~\ref{ovGS} reflect the 
situation of SU(2) symmetric interactions. The value for the 
interaction strengths defines the interval in $q$ where the 
system significantly occupies the $n=1$ level. For the relatively small 
value $g=0.2$ in Fig.~\ref{ovGS} (c), the drop in the overlap 
curve appears to be quite sharp. We can extend it over a larger 
region by choosing a stronger interaction. We have also investigated 
an interaction term, where $g_{\uparrow\uparrow}=g_{\downarrow\downarrow}$ 
and $g_{\uparrow\downarrow}=0$. Qualitatively, this yields the same 
results as in the SU(2)-symmetric case for $0<q^2<3B$.

The problem becomes analytically solvable if we
choose $g_{\uparrow\uparrow}=g_{\downarrow\downarrow}=0$ 
and $g_{\uparrow\downarrow}=g>0$ \cite{trombettoniPRA}. In this case, the
Laughlin state projected into the $n=0$ Landau level is a zero-energy
eigenstate of the two-body contact potential, and thus the true ground state
below the first degeneracy point. We have numerically confirmed
this prediction for $N=4$ particles. In particular we have checked that for
$\nu>1/2$ (that is for $N=4$ and $N_{\Phi}<8$), no states with zero interaction
energy solve the problem at finite spin-orbit coupling strength $q$.

\subsection{At the degeneracy point and slightly above: 
$3 \leq q^2/B \lesssim 3.2 $} 

Slightly above the degeneracy point (see Fig.~\ref{fig:ene}), when 
most particles populate the $n=1$ level, the projection of the 
Laughlin state according to Eq.~(\ref{p1}) represents reasonably well 
the true ground state. As seen in Fig.~\ref{ovGS} (a), 
also this projection yields a gapped state. As in the standard 
Laughlin state, the ground state is unique (apart from the 
center-of-mass degeneracy), and characterized by a Haldane 
momentum $\vec{K}=\vec{0}$. This gives another hint for the 
Laughlin-like behavior of the system.

The projected Laughlin states, however, fail to describe the 
system precisely at the degeneracy point. Here, the gap vanishes 
and a quasi-degenerate manifold of several states at different 
$\vec{K}$ describes the ground state. Apart from the fact 
that $\vec{K}=\vec{0}$ does not belong to this manifold, 
the lowest state at this pseudomomentum has no overlap with 
the projected Laughlin states. The failure of the Laughlin 
projections at the degeneracy point can be traced back to 
the fact that the projected Laughlin states are restricted 
to a single Landau level. We expect that at the degeneracy 
point, the particles have to be mapped partly according 
to Eq.~(\ref{p0}), and partly according to Eq.~(\ref{p1}), 
but we have no unique prescription for this. At the degeneracy 
point, our numerical results are very sensitive to particle 
number with respect to the number and Haldane momenta of 
the states within the ground state manifold.

\subsection{Clearly above the degeneracy point: 
$3.2 \lesssim q^2/B \lesssim 4$}

\begin {figure}[t]
\flushright
\includegraphics[width=0.8\textwidth]{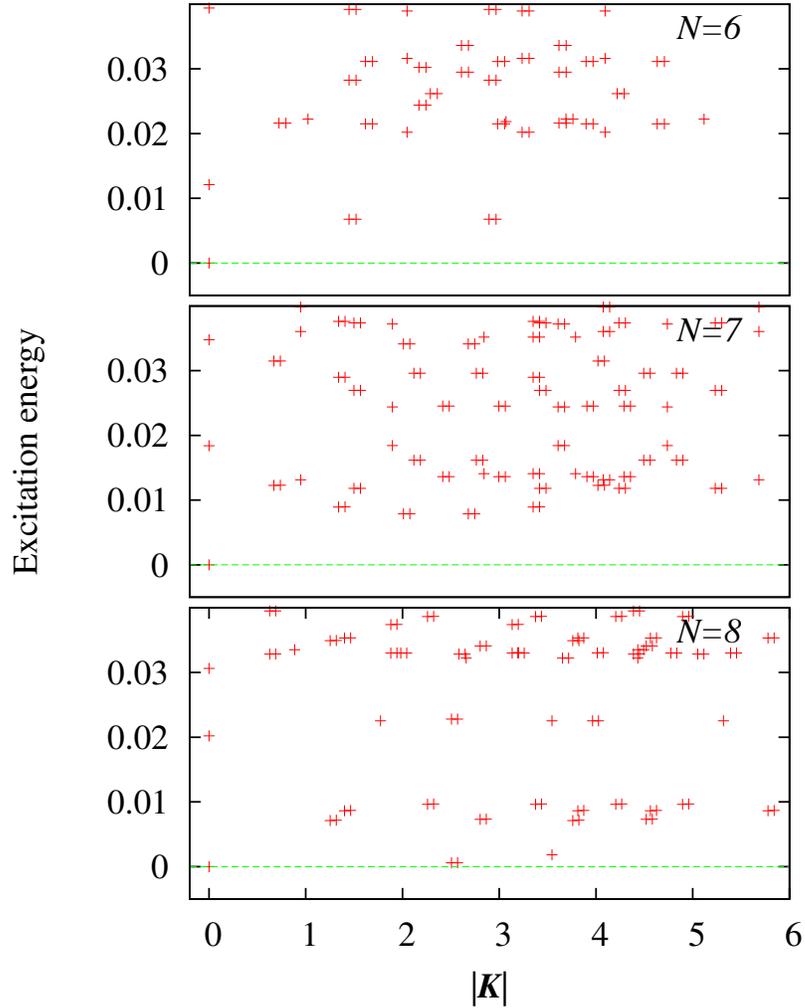}
\caption{\label{kspec} Excitation energies of low-lying 
states at $q^2/B=3.6$ for SU(2)-symmetric 
interactions ($g=0.2$). Energies are in units of $gB$, 
and $|\vec{K}|$ is in units of $2\pi \hbar/a$.}
\end {figure}

The regime where the $n=1$ Laughlin state projection describes 
well the system turns out to be restricted to a small parameter 
range above the degeneracy point. When further increasing the 
spin-orbit coupling strength, the overlap between true ground 
state and projected Laughlin state rapidly decreases and finally 
vanishes. For our choice of interaction of Fig.~\ref{ovGS} (c), 
SU(2) symmetric with $g=0.2$, it is zero already for $q^2/B =3.2$. 
Here, the system is still far away from the next degeneracy 
point at $q^2/B=5$, thus the two lowest Landau 
levels are still given by $n=1$ and $n=0$. Taking into account 
these two levels, we have performed the exact diagonalization 
for up to $N=6$ particles. Since in that region, roughly located 
around $3.2  \lesssim q^2/B \lesssim 4$, a significant gap 
separates the $n=1$ Landau level from the $n=0$ level, we also 
have performed exact diagonalization studies neglecting the $n=0$ 
level, which indeed has yielded the same results. 
With this reduction of the Hilbert space, we have reached system 
sizes of up to $N=8$ particles.

As before we have considered different interaction parameters. Now the
choice with $g_{\uparrow\uparrow}=g_{\downarrow\downarrow}=0$ 
and $g_{\uparrow\downarrow}>0$ does not lead to a ground state of zero
interaction energy in the region above the degeneracy point, since such a state
becomes unfavorable with respect to its single-particle energy.
The SU(2)-symmetric choice of interactions and  the configuration with
$g_{\uparrow\uparrow}=g_{\downarrow\downarrow}>0$ and
$g_{\uparrow\downarrow}=0$ yield qualitative differences in the spectral 
structure only for $N \leq 5$, which are absent in larger 
systems. We thus restrict ourselves to presenting the data obtained 
for the SU(2)-symmetric case. 

For all particle numbers, a state at $\vec{K}=\vec{0}$ is the 
ground state of the system, but this state is not well described 
by the projected Laughlin state. Neither is it protected by a 
significant gap: Depending on the particle number, states at 
different $\vec{K}$ are energetically very close to the ground 
state, see Fig.~\ref{kspec}. The upper spin component 
of this state is found to be well described by the ground 
state of a one-component system which is assumed to live 
in the first excited Landau level, see Fig.~\ref{ovGS} (c). This 
overlap further increases: For $N=8$ it reaches the value 
0.96, if we renormalize the upper spin component to one. 
Another state which has a sizable overlap with the true ground 
state is the ground state of a one-component system in the 
second excited Landau level. Comparing this state with the 
re-normalized lower spin component of the true ground state, 
the overlap is 0.38 for $N=6$ and decreases to 0.30 for $N=8$.

The relevance of these states is not very surprising, given 
the fact that the spin-orbit coupled level $\ket{n=1,j}$ is constructed 
from the first and second excited Landau level of the spinless 
problem. But it is noteworthy that the projection of the Laughlin 
state into these levels has basically no overlap with the 
respective ground states. Studying the behavior of particles in 
higher Landau levels is particularly relevant with respect 
to fermionic systems. In the field of strongly correlated 
electrons, very much attention has been paid to the Hall plateau 
at $\nu=5/2$. For a spin-polarized system, this represents the 
situation of a half-filled first excited Landau level, which 
might be described by a Moore-Read-like wave 
function~\cite{moore-read} and thus support non-Abelian excitations. 
With bosonic systems, it is not possible to study higher Landau 
levels by filling the lower levels. However, as our results 
clearly prove, the spin-orbit coupled system above the degeneracy 
point behaves with very high fidelity like a one-component 
system in the first excited Landau level. It thus becomes 
possible to explore this regime also within bosonic systems.

From the results shown in Fig.~\ref{kspec}, it is unclear 
whether the system develops a gapped phase in the thermodynamic 
limit, as the tendency observed for sizes up to $N=7$ may 
suggest. At $N=8$, however, a quasi-degeneracy of the 
$\vec{K}=\vec{0}$ state with several states at distinct 
Haldane momenta, indicates that a phase with a broken 
translational symmetry is established, like crystalline 
phases or bubble phases~\cite{fogler-bubble,haldane-bubble,fabian}. 
Such phases have been discussed in the context of fermions 
as possible candidates for substituting the Laughlin 
state in higher Landau levels.

\subsection{Laughlin states for $\nu<1/2$ \label{Q4laug}}
The Laughlin state defined in Eq. (\ref{laug}) is characterized by an integer
parameter $Q$, which turns out to be the inverse of the filling factor. So far
we have discussed only the most prominent case of $Q=2$, but any even value of
$Q$ yields a bosonic wave function of zero interaction energy. As has been
pointed out in Refs. \cite{trombettoni,trombettoniPRA}, for $Q>2$ this wave
function remains a zero interaction energy state when mapped from the lowest
Landau level of the system without spin-orbit coupling into the Landau level
structure of the spin-orbit coupled system according to Eq. (\ref{p0}). The
resulting wave function describes a system with all atoms exclusively in the
$\ket{n=0}$ level. It thus represents the ground state for spin-orbit coupling
strengths $q^2 \leq 3B$ at filling factor $\nu=1/Q$. 

We have explicitly checked for $Q=4$ (and $N=3, \ 4$) that such a zero
interaction energy state exists for $q^2 \leq 3B$. Also, no zero interaction
energy states are found for $q^2 \leq 3B$ at filling factors larger than
$\nu=1/4$. In agreement with its Laughlin-like character, it is located at
Haldane momentum $\vec{K}=(0,0)$, and it is the unique ground-state for $q^2 <
3B$. Precisely at the degeneracy point, $q^2 = 3B$, additional states with zero
interaction energy occur within a ground state manifold which then contains 7
(4) states for $N=4$ ($N=3$) at $\vec{K}=(0,0)$, and additional states at
different pseudomomenta.

The origin of one of the additional ground states at the degeneracy point can
be traced back to the $Q=4$ Laughlin state. Therefore one has to note that
every pair of coordinates appears with a $Q$th order zero of the form $z_i-z_j$
in the wave function of Eq. (\ref{laug}). Applying the mapping of Eq. (\ref{p0})
to a state contains operations where the original wave function is multiplied
with the complex conjugate variable $z_i^*$, and where the derivatives
$\partial_{z_i}$ are taken. Only the latter operation may affect the property
of being a zero interaction energy state. But since in every term $(z_i-z_j)^Q$
at most the second derivative $\partial_{z_i}\partial_{z_j}$ is taken, the
mapping of Eq. (\ref{p0}) applied to the wave function of Eq. (\ref{laug})
yields a wave function where every pair of particles still has a zero of at
least $(Q-2)$th order. For $Q=4$, this still allows to raise the Landau level
index of one particle from $n=0$ to $n=1$, yielding a partially raised $Q=4$
Laughlin state.

Indeed we find for $q^2<3B$ exactly one state with the total energy
$E=(N-1)E_0^{-}+E_1^-$. This state contains one particle in the $n=1$ level,
while the others are in $n=0$. It is thus seen to be a state
of zero interaction energy. As $E_1^->E_0^-$ below the degeneracy point, there
it appears as an excited state in the spectrum of the system. The symmetry of
this state is characterized by $\vec{K}=(0,0)$. By lowering the quantum number
of the particle in $n=1$ to $n=0$, we have been able to explicitly check that
this state is the partially raised Laughlin state.

The other zero interaction energy states occuring at the degeneracy point do
not exist in the spectrum for $q^2<3B$. They are not polarized
with respect to the Landau level quantum number, that is, not all contributing
Fock states contain the same number of particles in each Landau level. Only at
the degeneracy point such levels can be combined arbitrarily. This
property allows for a larger number of ground states at the degeneracy
point.

For larger spin-orbit coupling strengths, $q^2>3B$, the $Q=4$ Laughlin state and
its partially raised counterpart remain zero interaction energy eigenstates of
the system. However, their single-particle energy now is higher than the lowest
possible one given by $N E_1^-$. This will quickly give rise to a ground state
with higher occupation of the $n=1$ level. This state has finite interaction
energy, as predicted by Ref. \cite{trombettoniPRA}. The precise localization of
the transition into the new ground state depends on the interaction
strength. For the SU(2) symmetric choice with $g=1$, we get a ground state with
finite interaction energy already for $q^2=3.03 \sqrt{B}$.

A state of zero interaction energy which is fully polarized in the $n=1$
level is a Laughlin state with $Q=6$ projected into $n=1$. At such a low
filling, $\nu=1/6$, our numerics is restricted to $N=3$ particles. Of course, a
lot of states with zero interaction energy exist at this filling, but only one
has the property of being completely within the $n=1$ level. Such state is not
found at higher filling, that is for $N=3$ and $N<N_{\Phi}=18$. Obviously, this
state is the true ground state between the first and the second degeneracy
point. If we consider, instead of the SU(2) symmetric choice, an interaction
only between different pseudospins, that is
$g_{\uparrow\uparrow}=g_{\downarrow\downarrow}=0$ 
and $g_{\uparrow\downarrow}>0$, zero interaction energy states within
the $n=1$ level have been predicted for fillings up to $\nu=1/4$
\cite{trombettoniPRA}. We have checked this prediction for $N=4$. Dor
spin-orbit coupling strengths $q$ which are sufficiently far above the
degeneracy point, the unique ground state, with Haldane symmetry
$\vec{K}=(0,0)$, is given by this fully polarized state of zero interaction
energy.

\section{Incompressible phases at the first degeneracy point\label{secim}}

We have seen in the previous section that above and below the degeneracy point
the system behaves as an incompressible Laughlin liquid at $\nu=1/Q$ with $Q$
even. As we have shown in most detail for $\nu=1/2$, the incompressibility is
lost at the degeneracy point, where the number of states in the lowest Landau
level is doubled, giving rise to a degeneracy also on the many-body level. In
this section we investigate the system at this degeneracy point, and look for
filling factors which might support gapped phases. All results presented below
have been obtained for an SU(2) symmetric choice of interactions.

\subsection{$\nu=2$}

\begin {figure}
\flushright
\includegraphics[width=0.8\textwidth]{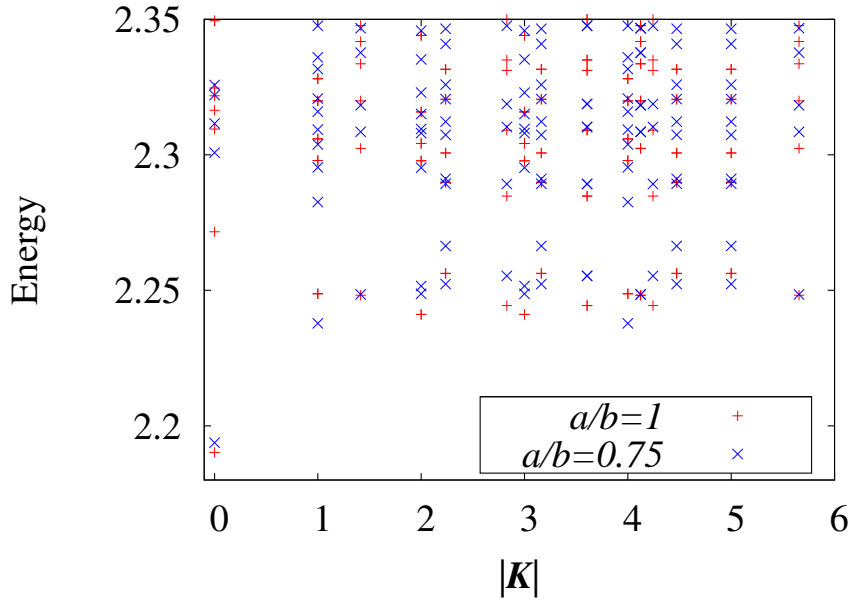}
\caption{\label{N10} Spectrum at $\nu=2$ for SU(2) symmetric interactions. Axis
ratios 1 and 0.75 are plotted. Note that the degeneracy between $90^{\circ}$
rotations in $\vec{K}$-space is lifted for ratios different from 1, and
therefore, more states seem to be present at ratio 0.75. Energies in units of
$gB$, and $|\vec{K}|$ in units of $2\pi \hbar/a$.}
\end {figure}

A clearly gapped phase shows up at filling $\nu=2$, where a unique 
ground state at $\vec{K}=\vec{0}$ is found for $N=8$ and $N=10$. 
This fact makes this phase distinct from the Read-Rezayi states and 
the NASS states which are both possible candidates due to the filling 
factor. Apart from the large gap, the robustness of the spectrum 
against deformations of the torus ratio, shown in Fig.~\ref{N10}, 
signals the incompressibility of this phase. 

\subsection{$\nu=3/2$}

As has been pointed out in several works \cite{trombettoni, palmer-pachos,
komineas-cooper}, off the degeneracy points incompressible phases are supported
not only at $\nu=1/2$, but also at other fillings corresponding to the
Read-Rezayi series \cite{read-rezayi}. 
At the Read-Rezayi filling, $\nu=3/2$ for $N=6$, a 
ground-state at $\vec{K}=\vec{0}$ is 
separated by a sizable gap of about twice the typical energy difference 
from a second state at $\vec{K}=\vec{0}$. Since a two-fold degenerate 
ground state at $\vec{K}=\vec{0}$ characterizes the $\nu=3/2$ Read-Rezayi 
state on the torus, one might expect that at this filling a Read-Rezayi 
phase exists on the degeneracy point. We have thus calculated the 
ground-states of a spin-polarized system with four-body contact interaction. 
Projecting the two zero-energy states into the Landau level structure of the 
spin-orbit coupled system, we obtain an overlap of 0.70 with the real ground
state. 
This number is close to the total weight of the fully polarized contribution to
this state with all atoms in the $n=0$ level, being 0.72. For the second state,
an overlap of 0.32 is found. Increasing the particle number to $N=9$, we 
still find overlaps of 0.39 and 0.42, but the spectral structure is not 
robust. While one state at $\vec{K}=\vec{0}$ remains the ground state, 
the second state at this pseudomomentum lies above other states in the spectrum.

\subsection{$\nu=1$}

In spin-polarized Bose systems, the famous Moore-Read state is supported at filling factor 
$\nu=1$. Also for spin-orbit coupled systems, this state plays a role: Below the degeneracy point ($q^2<3B$), the three lowest eigenstates are found at  $\vec{K}=(N_{\Phi}/2,0)$, $(0,N_{\Phi}/2)$, and $(N_{\Phi}/2, N_{\Phi}/2)$. The phase can thus be identified with the
Moore-Read phase \cite{palmer-pachos}. However, this spectral structure disappears at the degeneracy point. Here we
find ground states at different pseudomomenta, which even depend on the size of system: $(3,3)$, $(3,5)$, $(5,3)$, $(5,5)$, $(4,0)$, and $(0,4)$ for $N=5$, and $(3,0)$, $(0,3)$, $(2,2)$, $(2,4)$, $(4,2)$, and $(4,4)$ for $N=6$. For $N=7$, states at $(2,0)$, $(5,0)$, $(0,2)$, and $(0,5)$ provide the ground state manifold. For $N=8$, we have $(3,3)$, $(3,5)$, $(5,3)$, $(5,5)$, $(4,0)$, and $(0,4)$. Since the gap between these states, compared to other energy differences in the spectrum, is not extraordinarily big, the system is expected to be compressible.

\subsection{NASS series}

\begin {figure}[t]
\flushright
\includegraphics[width=1.\textwidth]{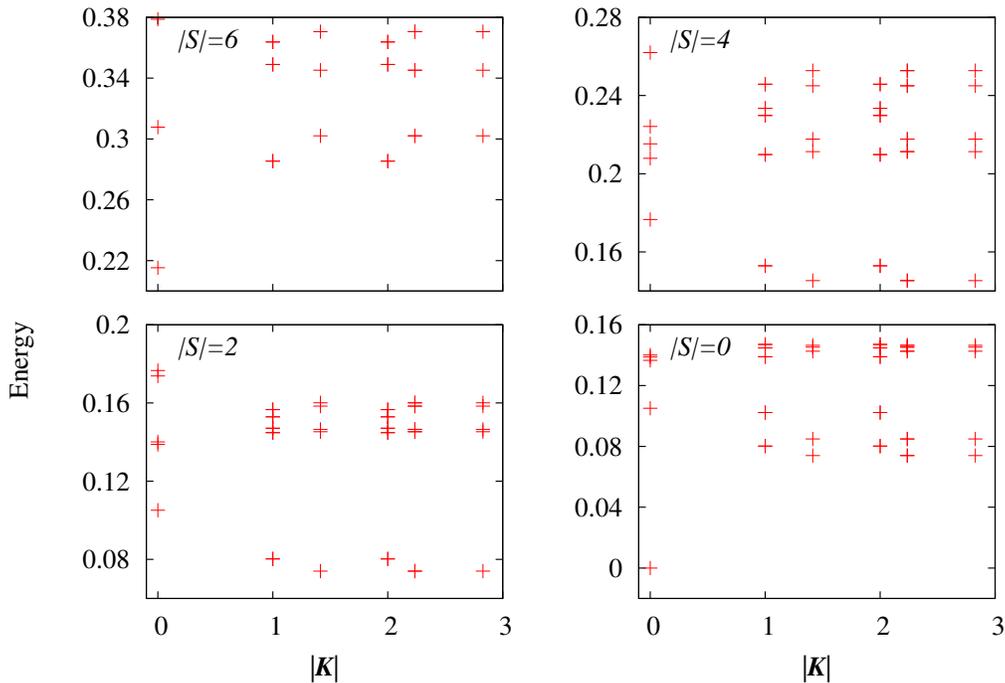}
\caption{\label{x2} Spectra of a two-component Bose gas without spin-orbit
coupling ($q=0$) at filling factor $\nu=2/3$ for different spin
polarizations $S$. $N=6$.}
\end {figure}

\begin {figure}[t]
\flushright
\includegraphics[width=0.82\textwidth]{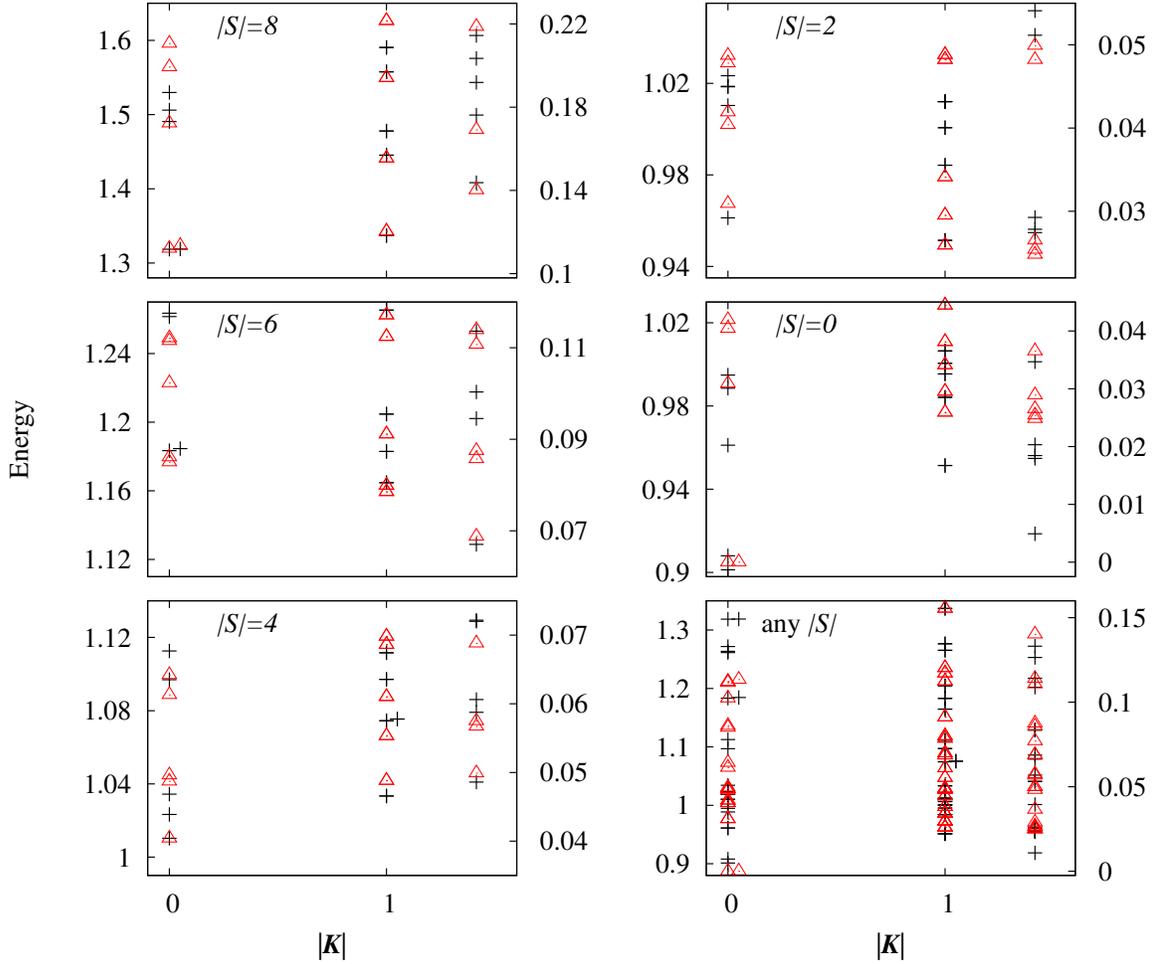}
\caption{\label{x3} Spectra of a two-component Bose gas without spin-orbit
coupling ($q=0$) at filling factor $\nu=4/3$ for different spin
polarizations $S$. The results for a two-body contact interaction 
are given by the black crosses which correspond to the energy scale 
on the left. The red triangles with the energy scale on the right 
show the results for a system interacting via a three-body contact 
potential.}
\end {figure}

The NASS filling factors, $\nu=2/3$ and $\nu=4/3$, have been recently 
shown to support incompressible phases in the system without 
spin-orbit coupling~\cite{rapido,furukawa}. It is instructive 
to recall the results in that limit, $q=0$, for our subsequent discussion 
at the first degenerate point. In Fig.~\ref{x2} we depict the 
spectrum of a system of six atoms interacting through a two-body 
contact interaction at $q=0$ at filling $\nu=2/3$. While for 
$|S|=6$ and $|S|=0$ a state with $\vec{K}=\vec{0}$ is the unique 
ground state, a (quasi)degenerate manifold of states with 
$\vec{K} \neq \vec{0}$ provides the ground state for spin 
polarizations $|S|=2$ and $|S|=4$. Interestingly, the 
energy of low-lying excitations at $|S|$ precisely agree with 
the ground state energy at $|S'|=|S|+2$. 

Similarly, for the $\nu=4/3$ filling, we have compared the 
spectrum obtained for a two-body contact interaction with the 
one obtained with only three-body contact interactions. Both spectra 
are depicted in Fig.~\ref{x3}.  The experimentally feasible case of two-body contact
interactions shares most of the spectral properties found in the three-body 
contact case in all polarizations. In particular, it is especially telling to note that 
the degeneracies of the lowest states for each polarization are shared 
in both calculations, a signal of the topological equivalence between 
both systems~\cite{wen-niu}. For $S=0$, the gapped ground state of the three-body contact interaction defines a NASS phase. 

\begin{figure}
\flushright
\includegraphics[width=0.98\textwidth]{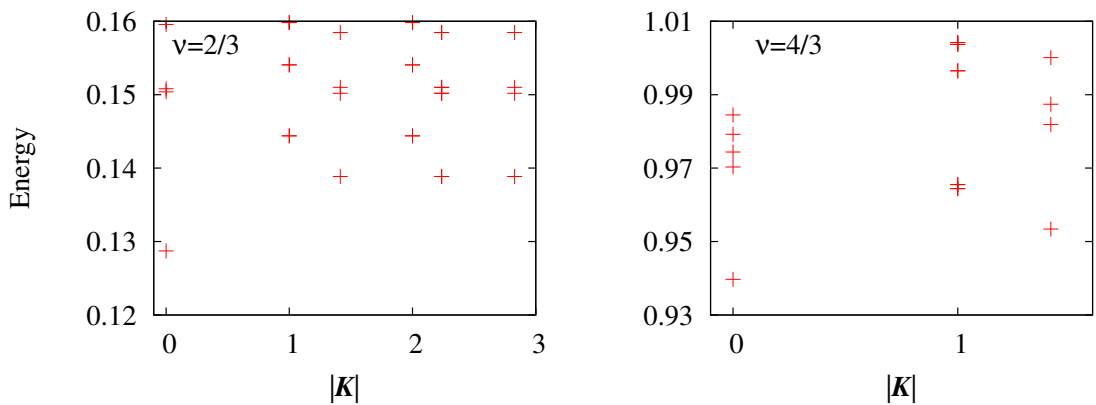}
\caption{\label{x1} Spectrum at the degeneracy point
$q^2/B=3$ for filling factor $\nu=2/3$ with $N=6$ (left) and for $\nu=4/3$ with
$N=8$ (right). We have chosen SU(2) symmetric interactions. Energies in
units of $gB$, and $|\vec{K}|$ in units of $2\pi \hbar/a$.}
\end {figure}

Now, we turn back to the spin-orbit coupled system. Here we do find 
a clear gap in the spectrum at $\nu=2/3$ with $N=6$, and a small gap 
at $\nu=4/3$ with $N=8$, see Fig.~\ref{x1}. 

For the $\nu=2/3$ filling, the nature of the ground state of 
the system can be related to the 221-Halperin state, Eq.~(\ref{221-halperin}).  
We have projected the 221-Halperin state, 
onto the Landau level structure of the spin-orbit coupled 
system according to Eq.~(\ref{p0}), and computed the overlap 
of this state with the $\nu=2/3$ ground state. We find the 
value of 0.18. It should be noted, however, that interactions now
can induce ``spin'' flips within the two-fold degenerate manifold. 
Therefore the analog of spin polarization, namely population 
imbalance $p\equiv N_0-N_1$, is not conserved. Here, $N_1$ ($N_0$) 
define the number of particles in the $n=1$ ($n=0$) Landau level. 
We therefore write the true ground state $\ket{\Psi}$ as
\begin{eqnarray}
\label{decomp}
\ket{\Psi} = \sum_{i=-N}^N w_i \ket{p=i}_{\Psi},
\end{eqnarray}
with $\ket{p=i}_{\Psi}$ the contributions with well-defined 
population imbalance. With a proper choice for the phase of each 
$\ket{p=i}_{\Psi}$, we can make $w_i$ a real number, measuring the weight of
each polarization $p$. Mapping the Halperin state 
with spin polarization $S=0$ to the spin-orbit coupled Landau levels according
to Eq. (\ref{p0})
yields a state with $p=0$. The overlap of this state with $\ket{\Psi}$ thus is
bounded by the corresponding weight,
$w_0=0.26$. Contrarily, the overlap of the projected Halperin state with
$\ket{p=0}_{\Psi}$ takes a significantly increased value, $0.18/0.26 \approx
0.7$.

Interestingly, the most important contributions to $\ket{\Psi}$ are
$\ket{p=6}_{\Psi}$ with a weight $w_{6}= 0.64$, $\ket{p=4}_{\Psi}$ with a
weight $w_{4}= 0.57$, and $\ket{p=2}_{\Psi}$ with a weight $w_{2}= 0.42$.
The contributions with $p<0$ are rarely populated with weights $w_{-2}=0.1$,
$w_{-4}=0.07$, and $w_{-6}=0.01$. This shows that population of $n=0$ is
still favored, despite the degeneracy with $n=1$ on the single-particle level.
It reflects a fact which we have already encountered before when analyzing
the system at $\nu=1/2$: Contact interactions favor the Landau level with
lower $n$.

As mentioned above, with good overlap we can interpret the $p=0$ contribution
as a projected Halperin state, that is, as the ground state of a
two-component system without spin-orbit coupling ($q=0$) in a two-body contact
potential with spin polarization $S=0$ and filling factor $\nu=2/3$.
Diagonalizing such a system for different $S$, and projecting the lowest state
with $\vec{K}=\vec{0}$ according to Eq.~(\ref{p0}), we are able to relate all
the states $\ket{p=i}_\Psi$ in the decomposition of Eq.~(\ref{decomp}) to ground
states of the two-body contact potential for a fixed spin polarization $S$ and
Haldane momentum $\vec{K}$. Explicitly, we find an overlap of 0.94 between the
projected ground state at $S=6$ and $\ket{p=6}_{\Psi}$, an overlap of 0.84 for
$S=4$, and an overlap of 0.83 for $S=2$. Modeling a state with the projected
ground states for all spin polarizations $-6 \leq S \leq 6$, we are able to
reproduce the true ground state $\ket{\Psi}$ with a fidelity of 0.82.

For $\nu=4/3$ with $N=8$, the overlap with the projected NASS state is 
only 0.03. Nevertheless, the true ground state can be reproduced from 
the lowest $\vec{K}=\vec{0}$ states of a three-body contact interaction, 
if we take into account different spin polarizations $S$ and project 
them according to Eq.~(\ref{p0}). With this we are able to model the 
true ground state of the spin-orbit coupled system with a total 
fidelity of 0.52.

\subsection{Low filling}

Finally we note that an incompressible phase can be expected at 
a critical $\nu_0$, defined as the largest filling supporting states 
with zero interaction energy. For any $\nu<\nu_0$, the interaction 
energy should remain zero, while for $\nu>\nu_0$ it is by definition 
larger than zero. Thus, at $\nu_0$ a kink in the energy as a 
function of $\nu$ can be expected, resulting in an incompressibility 
of the system. As Fig.~\ref{ovGS} (b) shows, the spin-orbit coupled 
system with two-body contact interaction has a finite interaction 
energy at filling $\nu=1/2$, in contrast to the system with only 
an Abelian gauge field. This is a consequence of the fact that, 
at this filling, there are no zero-energy states of two-body 
contact interaction within the first excited or higher
Landau levels. We thus should focus on smaller filling factors.

By analytical arguments, a zero interaction energy state of the form of a
(464) Halperin state has been predicted in Ref.~\cite{trombettoniPRA}.
With half of the particles in the $n=0$ level, and half of the particles in the
$n=1$ level, this state has been constructed as an unpolarized state with $p=0$.
However, such a state is ill-defined on a torus: The filling factor of the
$n=0$ level is given as $\nu_{n=0}=1/(4+4)=1/8$, while for $n=1$ it reads
$\nu_{n=1}=1/(4+6)=1/10$. Thus, the state cannot have $p=0$, if both levels are
assumed to fill the same area.

However, for arbitrary $p$, we are able to find states of zero interaction
energy at even larger filling factors. As already mentioned in Sec.
\ref{Q4laug}, also $\nu=1/4$ supports states with
zero interaction energy. Apart from the Laughlin state ($p=N$) and a Laughlin
state with one particle shifted into the first excited Landau level ($p=N-2$),
which are present in the spectrum of the system at any $q$, additional
zero-interaction appear at the degeneracy point. For $N=4$ ($N=3$), there are 5
(2) additional zero interaction energy states with $K=(0,0)$. Further states
with zero interaction energy appear at finite pseudomomenta. For $N=3$, there
are two such states for each pseudomomenta, while for $N=4$ this number varies
between 4 and 5. These zero interaction energy states do not exist for $q^2<3B$.

The huge degeneracy of zero interaction energy eigenstates at $\nu=1/4$ suggests
that this is not yet the critical filling $\nu_0$ at the degeneracy point. In
fact, we find a unique zero interaction energy state at $\vec{K}=(0,0)$ for
$\nu=2/7$ and $N=4$. This state has no well defined polarization $p$, but the
large weight $w_4=0.9298$ indicates that the atoms are mostly in the lowest
Landau level, $n=0$. However, this makes the state to be an eigenstate only at
the degeneracy point, leaving the Laughlin filling $\nu=1/4$ to be the
critical filling off the degeneracy point. We do not find states
with zero interaction energy are for $\nu=1/3$ (for $N=4$ and $N=5$) or
$\nu=4/13$ (for $N=4$). Here, we therefore expect $\nu=2/7$ to be the critical
filling of the system.

This would shift the incompressible phase from $\nu=1/4$ for $q^2<3B$ to
$\nu=2/7$ for $q^2=3B$. However, we should also note that the excitation gap
above the ground state at $\nu=2/7$ state takes the tiny value of $2 \cdot
10^{-4} gB$. The energy as a function of $\nu$ is a concave function, and in the
thermodynamic limit it might happen that $\frac{{\rm d}E}{{\rm d}\nu}
\rightarrow 0$, when $\nu \searrow \nu_0$. This would render a compressible 
phase even at $\nu_0$. We therefore conclude that, at 
the degeneracy point, robust incompressible phases are rather 
located in the denser regime. This should facilitate the 
experimental observation.

\section{Summary and conclusions\label{sum}}

We have performed a numerical study of the ground and excited states 
of an ultracold pseudospin 1/2 bosonic cloud subjected to an artificial 
magnetic field. The setup considered, first proposed in 
Ref.~\cite{trombettoni}, includes a spin-orbit coupling, which 
is mathematically equivalent to a non-Abelian SU(2) gauge potential. 
The single particle spectrum has a number of degenerate points 
which have received a special attention, as two different 
Landau levels simultaneously provide the low-energy manifold of single-particle
states. This fact makes them particularly appealing for exploring
interesting new phases.  

We have concentrated on the parameter region around the first 
degenerate point $q^2/B=3$. First, we have analyzed $\nu=1/2$ and 
discussed the different Laughlin-like states which have a large 
overlap with the ground state of the system. The exact ground 
state of the system at this filling is described either 
by the usual Laughlin wave function (mapped into our Landau level structure), 
if we are below the degenerate point, or by the corresponding 
ground state of contact interactions in the first excited Landau level, for
values of $q$ above the degenerate point. This allows to explore higher Landau
levels even with bosons. Slightly above 
the degenerate point there is a sizable overlap between the 
ground state and a Laughlin state projected into the first excited Landau level.
At filling $\nu=1/4$, we find two Laughlin-like states with zero
interaction energy even in the spin-orbit coupled system. One of them is the
ground state for $q^2/B<3$. Additional states of zero interaction energy are
found at the degeneracy point for $\nu=1/4$, and it turns out that the
critical filling is shifted to $\nu=2/7$. Above the degeneracy point, ground
states with zero interaction energy are found only for $\nu=1/6$. An exception
are configurations without intraspecies interactions.

Secondly, we have studied in detail the phases appearing 
at the first degenerate point for the most interesting 
filling factors. Notably, we find an unexpected gapped phase 
at $\vec{K}=\vec{0}$ at a fairly dense filling, $\nu=2$. This phase 
is gapped for $N=8$ and $N=10$ and is robust for changes in 
the aspect ratio of the torus, as expected for an incompressible 
phase. Motivated by the $\nu=1/2$ results, we looked for 
incompressible phases at other Read-Rezayi fillings, $\nu=1$ and 
$\nu=3/2$. For $\nu=1$ we find no indication of a gap in the spectrum 
of the system. For the $\nu=3/2$ we find a sizable overlap between 
the exact ground state of the system and the ground state obtained 
with four-body contact interactions mapped into our Landau level 
structure. The spectral structure however is not robust as we 
increase the particle number, finding that the two-fold degeneracy 
of the $\vec{K}=\vec{0}$ ground state is lifted when we go from 
six to nine particles. 

Thirdly, we have analyzed the gapped phases occuring at the NASS filling factors $\nu=2/3$ and $\nu=4/3$. Decomposing the ground states into the contributions with a fixed population imbalance between the two Landau levels, we are able to relate these states to the ground states of a two-body contact potential for $\nu=2/3$, or of a three-body contact potential for $\nu=4/3$, obtained for fixed spin polarizations in a system without spin-orbit coupling.

As a conclusion, we note that the properties of the spin-orbit coupled system in
the vicinity of the degeneracy point make the considered setup very appealing
from the experimental point of view: Upon tuning the spin-orbit coupling
strength, transitions between different topological phases can be realized. It
also becomes possible to explore the behavior of bosons in higher Landau levels.
For the relatively dense system at filling $\nu=2$, a robust gapped phase occurs
which is neither described by the NASS nor the Read-Rezayi series.

\ack
The authors thank Nuria Barber\'an for interesting discussions on the ``fate of
the Laughlin'' state, Sec.~\ref{laugfate}. 
This work has been supported by EU (NAMEQUAM, AQUTE), 
ERC (QUAGATUA), Spanish MINCIN (FIS2008-00784 TOQATA), Alexander von Humboldt
Stiftung, and AAII-Hubbard. B.~J.-D. is supported by the Ram\'on y Cajal
program. M. L. acknowledges support from the Joachim Herz Foundation and Hamburg
University.

\appendix
\section{Interaction matrix elements}
\label{appA}

The interaction matrix elements $V_{\{n,j\}}$, with $\{n,j\}$ 
denoting the set of quantum numbers $n_1,\dots,n_4$ 
and $j_1,\dots,j_4$, are given by
\begin{eqnarray}
 V_{\{ n,j \} } = \bra{n_{1},j_{1}} \bra{n_{2},j_{2}}  V \ket{n_{3},j_{3}} \ket{n_{4},j_{3}},
\end{eqnarray}
with $V$ given in Eq.~(\ref{V}). This yields:
\begin{eqnarray}
\label{Vmunu}
 V_{\{n,j\}} &=& \frac{1}{2} \Big( \
g_{\uparrow\uparrow} \ \alpha_{n_1}^- \alpha_{n_2}^- \alpha_{n_3}^- \alpha_{n_4}^-\ {\cal I}_{n_1,n_2,n_3,n_4}^{j_1,j_2,j_3,j_4} \nonumber \\ 
& +& 
g_{\downarrow\downarrow} \ \beta_{n_1}^- \beta_{n_2}^- \beta_{n_3}^- \beta_{n_4}^-\ {\cal I}_{n_1+1,n_2+1,n_3+1,n_4+1}^{j_1,j_2,j_3,j_4} \nonumber \\ 
& +&
g_{\uparrow\downarrow} \ \alpha_{n_1}^- \beta_{n_2}^- \alpha_{n_3}^-\beta_{n_4}^- \ {\cal I}_{n_1,n_2+1,n_3,n_4+1}^{j_1,j_2,j_3,j_4} \nonumber \\ 
& +&
g_{\downarrow\uparrow} \ \beta_{n_1}^- \alpha_{n_2}^- \beta_{n_3}^-\alpha_{n_4}^- \ {\cal I}_{n_1+1,n_2,n_3+1,n_4}^{j_1,j_2,j_3,j_4} \
\Big),
\end{eqnarray}
with 
\begin{eqnarray}
\label{V6}
{\cal I}_{\{n\}}^{\{j\}} &= & 
\int_{0}^{a} \int_{0}^{b} \mathrm{d}^2 \vec{r}_1 \mathrm{d}^2 \vec{r}_2 \ 
\delta(\vec{r}_1-\vec{r}_2) \Psi^*_{n_1j_1}(\vec{r}_1) \Psi^*_{n_2j_2}(\vec{r}_2) \nonumber \\ 
& \times& \Psi_{n_3j_3}(\vec{r}_1) \Psi_{n_4j_4}(\vec{r}_2).
\end{eqnarray}
To evaluate this integral, we write the elliptic $\theta$ 
functions in $\Psi_{nj}$ as infinite sums, see also~\cite{chakraborty, karel}. 
We get
\begin{eqnarray}
\label{stsums}
{\cal I}_{\{n\}}^{\{j\}} &=& \frac{1}{2\pi N_{\Phi}} 
\sum_{\mu=-\infty}^{\infty} \sum_{\nu=-\infty}^{\infty} 
\delta'_{j_1+j_2,j_3+j_4} \delta'_{j_1-j_4,t}\nonumber \\ & \times&
C_{n_1,n_4}\left(\frac{\tau}{\vartheta}\mu, \tau \vartheta \nu \right)
C_{n_2,n_3} \left(-\frac{\tau}{\vartheta}\mu, -\tau \vartheta \nu \right)
\nonumber \\ & \times&
 \exp\left[ -\tau^2 \left\{ \frac{1}{2} \left( \left(\frac{\mu}{\vartheta}\right)^2 
+ \left(\nu \vartheta \right)^2  \right)  + i \mu (j_1-j_3) \right\} \right] \,.
\end{eqnarray}
where $\delta'$ is a Kronecker delta modulo $N_{\Phi}$, $\tau=\sqrt{2\pi/N_{\Phi}}$, 
and $\vartheta=\sqrt{a/b}$. The coefficients $C_{n_i,n_j}$ read 
(with $u,v \in \mathbb{R}$):
\begin{eqnarray*}
 C_{1,0}(u,v) &=& -(u+iv)/\sqrt{2} = -C_{0,1}^*(u,v),\\
 C_{1,1}(u,v) &=& 1 + C_{1,0}(u,v) C_{0,1}(u,v),\\
 C_{2,0}(u,v) &=& \sqrt{2}[C_{0,1}(u,v)]^2 = -C_{0,2}^*(u,v),\\
 C_{1,2}(u,v) &\equiv& (u^2+v^2-4)(u+i v)/4 = -C_{2,1}^*(u,v),\\
 C_{2,2}(u,v) &\equiv &(8-8(u^2+v^2)+2u^2v^2 +u^4+v^4)/8.
\end{eqnarray*}
The sums in Eq.~(\ref{stsums}) converge quickly, especially within 
the lowest Landau level. For levels up to $n=2$, taking into 
account only values with $|\mu| \leq 16$ and $|\nu| \leq 8$ provides 
very precise values. 

\clearpage
\bibliographystyle{iopart-num}

\begin{thebibliography}{10}
\expandafter\ifx\csname url\endcsname\relax
  \def\url#1{{\tt #1}}\fi
\expandafter\ifx\csname urlprefix\endcsname\relax\def\urlprefix{URL }\fi
\providecommand{\eprint}[2][]{\url{#2}}

\bibitem{laughlin}
Laughlin R~B 1983 {\em Phys. Rev. Lett.\/} {\bf 50} 1395

\bibitem{arovas}
Arovas D, Schrieffer J~R and Wilczek F 1984 {\em Phys. Rev. Lett.\/} {\bf
  53}(7) 722--723

\bibitem{moore-read}
Moore G and Read N 1991 {\em Nucl. Phys. B\/} {\bf 360} 362

\bibitem{wen-niu}
Wen X~G and Niu Q 1990 {\em Phys. Rev. B\/} {\bf 41}(13) 9377--9396

\bibitem{cooper-aip}
Cooper N 2008 {\em Adv. Phys.\/} {\bf 57} 539

\bibitem{fetter-revmod}
Fetter A~L 2009 {\em Rev. Mod. Phys.\/} {\bf 81} 647--691

\bibitem{cooperwilkin}
Cooper N~R and Wilkin N~K 1999 {\em Phys. Rev. B\/} {\bf 60}(24) R16279--R16282

\bibitem{wilkingunn}
Wilkin N~K and Gunn J~M~F 2000 {\em Phys. Rev. Lett.\/} {\bf 84}(1) 6--9

\bibitem{cooper-wilkin-gunn}
Cooper N~R, Wilkin N~K and Gunn J~M~F 2001 {\em Phys. Rev. Lett.\/} {\bf
  87}(12) 120405

\bibitem{schweikhard}
Schweikhard V, Coddington I, Engels P, Mogendorff V~P and Cornell E~A 2004 {\em
  Phys. Rev. Lett.\/} {\bf 92} 040404

\bibitem{cornell-vortices}
Matthews M~R, Anderson B~P, Haljan P~C, Hall D~S, Wieman C~E and Cornell E~A
  1999 {\em Phys. Rev. Lett.\/} {\bf 83}(13) 2498--2501

\bibitem{cornell-vortex-lattice}
Schweikhard V, Coddington I, Engels P, Tung S and Cornell E~A 2004 {\em Phys.
  Rev. Lett.\/} {\bf 93}(21) 210403

\bibitem{dalibard-vortex2002}
Rosenbusch P, Petrov D~S, Sinha S, Chevy F, Bretin V, Castin Y, Shlyapnikov G
  and Dalibard J 2002 {\em Phys. Rev. Lett.\/} {\bf 88}(25) 250403

\bibitem{dalibard-vortex2004}
Bretin V, Stock S, Seurin Y and Dalibard J 2004 {\em Phys. Rev. Lett.\/} {\bf
  92}(5) 050403

\bibitem{dalibard}
Dalibard J, Gerbier F, {Juzeli\ifmmode \bar{u}\else \={u}\fi{}nas} G and
  {\"O}hberg P 2011 {\em Rev. Mod. Phys.\/} {\bf 83}(4) 1523--1543

\bibitem{brunoPRA}
Juli{\'a}-D{\'i}az B, Dagnino D, G{\"u}nter K~J, Gra{\ss} T, Barber{\'a}n N,
  Lewenstein M and Dalibard J 2011 {\em Phys. Rev. A\/} {\bf 84}(5) 053605

\bibitem{bruno-njp}
Juli{\'a}-D{\'i}az B, Gra{\ss} T, Barber{\'a}n N and Lewenstein M 2012 {\em New
  J. Phys.\/} {\bf 14} 055003

\bibitem{spielmanPRL}
Lin Y~J, Compton R~L, Perry A~R, Phillips W~D, Porto J~V and Spielman I~B 2009
  {\em Phys. Rev. Lett.\/} {\bf 102} 130401

\bibitem{lin}
Lin Y~J, Compton R~L, Jim{\'e}nez-Garc{\'i}a K, Porto J~V and Spielman I~B 2009
  {\em Nature\/} {\bf 462} 628

\bibitem{spielman-peierls}
Jim{\'e}nez-Garc{\'i}a K, LeBlanc L~J, Williams R~A, Beeler M~C, Perry A~R and
  Spielman I~B 2012 {\em Phys. Rev. Lett.\/} {\bf 108}(22) 225303

\bibitem{bloch-gauge}
Aidelsburger M, Atala M, Nascimb{\`e}ne S, Trotzky S, Chen Y~A and Bloch I 2011
  {\em Phys. Rev. Lett.\/} {\bf 107}(25) 255301

\bibitem{sengstock12}
Struck J, {\"O}lschl{\"a}ger C, Weinberg M, Hauke P, Simonet J, Eckardt A,
  Lewenstein M, Sengstock K and Windpassinger P 2012 {\em Phys. Rev. Lett.\/}
  {\bf 108}(22) 225304

\bibitem{rashba}
Bychkov Y~A and Rashba E~I 1984 {\em J. Phys. C\/} {\bf 17} 6039

\bibitem{juzeli2004}
{Juzeli\ifmmode \bar{u}\else \={u}\fi{}nas} G and {\"O}hberg P 2004 {\em Phys.
  Rev. Lett.\/} {\bf 93} 033602

\bibitem{fleischhauer}
Ruseckas J, {Juzeli\ifmmode \bar{u}\else \={u}\fi{}nas} G, {\"O}hberg P and
  Fleischhauer M 2005 {\em Phys. Rev. Lett.\/} {\bf 95} 010404

\bibitem{galitski}
Stanescu T~D, Anderson B and Galitski V 2008 {\em Phys. Rev. A\/} {\bf 78}(2)
  023616

\bibitem{spielman-sobec}
Lin Y~J, Jim{\'e}nez-Garc{\'i}a K and Spielman I~B 2011 {\em Nature\/} {\bf
  471} 83--86

\bibitem{williams12}
Williams R~A, LeBlanc L~J, Jim{\'e}nez-Garc{\'i}a K, Beeler M~C, Perry A~R,
  Phillips W~D and Spielman I~B 2012 {\em Science\/} {\bf 335} 314--317

\bibitem{sinha11}
Sinha S, Nath R and Santos L 2011 {\em Phys. Rev. Lett.\/} {\bf 107}(27) 270401

\bibitem{hu12}
Hu H, Ramachandhran B, Pu H and Liu X~J 2012 {\em Phys. Rev. Lett.\/} {\bf
  108}(1) 010402

\bibitem{ryan}
Barnett R, Powell S, Gra{\ss} T, Lewenstein M and {Das Sarma} S 2012 {\em Phys.
  Rev. A\/} {\bf 85}(2) 023615

\bibitem{ryan-erratum}
Barnett R, Powell S, Gra{\ss} T, Lewenstein M and {Das Sarma} S 2012 {\em Phys.
  Rev. A\/} {\bf 85}(4) 049905

\bibitem{ozawa-baym-stability}
Ozawa T and Baym G 2012 {\em Phys. Rev. Lett.\/} {\bf 109}(2) 025301

\bibitem{ozawa-baym-stripes}
Ozawa T and Baym G 2012 {\em Phys. Rev. A\/} {\bf 85}(6) 063623

\bibitem{SOstringari}
Li Y, Pitaevskii L~P and Stringari S 2012 {\em Phys. Rev. Lett.\/} {\bf
  108}(22) 225301

\bibitem{osterloh}
Osterloh K, Baig M, Santos L, Zoller P and Lewenstein M 2005 {\em Phys. Rev.
  Lett.\/} {\bf 95} 010403

\bibitem{philipna}
Hauke P, Tieleman O, Celi A, {\"O}lschl{\"a}ger C, Simonet J, Struck J,
  Weinberg M, Windpassinger P, Sengstock K, Lewenstein M and Eckardt A 2012
  {\em Phys. Rev. Lett.\/} {\bf 109}(14) 145301

\bibitem{indianpra}
Gra{\ss} T, Saha K, Sengupta K and Lewenstein M 2011 {\em Phys. Rev. A\/} {\bf
  84}(5) 053632

\bibitem{trivediBH}
Cole W~S, Zhang S, Paramekanti A and Trivedi N 2012 {\em Phys. Rev. Lett.\/}
  {\bf 109}(8) 085302

\bibitem{galitski12}
{Radi\ifmmode \acute{c}\else {\'c}\fi{}} J, {Di Ciolo} A, Sun K and Galitski V
  2012 {\em Phys. Rev. Lett.\/} {\bf 109}(8) 085303

\bibitem{fqhe-gap}
Gra{\ss} T, Baranov M~A and Lewenstein M 2011 {\em Phys. Rev. A\/} {\bf 84}(4)
  043605

\bibitem{rapido}
Gra{\ss} T, Juli{\'a}-D{\'i}az B, Barber{\'a}n N and Lewenstein M 2012 {\em
  Phys. Rev. A\/} {\bf 86}(2) 021603

\bibitem{furukawa}
Furukawa S and Ueda M 2012 {\em Phys. Rev. A\/} {\bf 86}(3) 031604

\bibitem{ardonne-schoutens}
Ardonne E and Schoutens K 1999 {\em Phys. Rev. Lett.\/} {\bf 82}(25) 5096--5099

\bibitem{nass-nucl}
Ardonne E, Read N, Rezayi E and Schoutens K 2001 {\em Nucl. Phys. B\/} {\bf
  607} 549--576

\bibitem{trombettoni}
Burrello M and Trombettoni A 2010 {\em Phys. Rev. Lett.\/} {\bf 105} 125304

\bibitem{trombettoniPRA}
Burrello M and Trombettoni A 2011 {\em Phys. Rev. A\/} {\bf 84}(4) 043625

\bibitem{read-rezayi}
Read N and Rezayi E 1999 {\em Phys. Rev. B\/} {\bf 59}(12) 8084--8092

\bibitem{palmer-pachos}
Palmer R~N and Pachos J~K 2011 {\em New J. Phys.\/} {\bf 13} 065002

\bibitem{komineas-cooper}
Komineas S and Cooper N~R 2012 {\em Phys. Rev. A\/} {\bf 85}(5) 053623

\bibitem{tripod-interactions}
Zhang Y, Mao L and Zhang C 2012 {\em Phys. Rev. Lett.\/} {\bf 108}(3) 035302

\bibitem{halperin}
Halperin B~I 1983 {\em Helv. Phys. Acta\/} {\bf 56} 75

\bibitem{yoshioka-halperin-lee}
Yoshioka D, Halperin B~I and Lee P~A 1983 {\em Phys. Rev. Lett.\/} {\bf 50}(16)
  1219--1222

\bibitem{haldane}
Haldane F~D~M 1985 {\em Phys. Rev. Lett.\/} {\bf 55}(20) 2095--2098

\bibitem{chakraborty}
Chakraborty T and Pietilainen P 1995 {\em The Quantum Hall Effects\/}
  (Springer)

\bibitem{fogler-bubble}
Fogler M~M and Koulakov A~A 1997 {\em Phys. Rev. B\/} {\bf 55}(15) 9326--9329

\bibitem{haldane-bubble}
Haldane F~D~M, Rezayi E~H and Yang K 2000 {\em Phys. Rev. Lett.\/} {\bf 85}(25)
  5396--5399

\bibitem{fabian}
Grusdt F and Fleischhauer M 2012 {\em arXiv:1207.3716\/}

\bibitem{karel}
V{\'y}born{\'y} K 2005 {\em Spin in fractional quantum Hall systems\/} Ph.D.
  thesis University of Hamburg

\end{thebibliography}
\providecommand{\newblock}{}

\end{document}